\documentclass[11pt]{article}
\usepackage{jheppubmodessay}
\pdfoutput=1
\usepackage[showonlyrefs]{mathtools}
\usepackage{psfrag}
\usepackage{array}
\usepackage{amssymb}
\usepackage{graphicx}
\usepackage[labelsep=quad]{subcaption}
\mathtoolsset{showonlyrefs}
\usepackage{epstopdf}
	
\usepackage{epsfig}
\usepackage[punctsep]{collref}
\collectsep[]{;}

\def\al{A}

\def\Or[#1]{{\text{O}}\left({#1}\right)}
\def\dotl[#1,#2]{\left\langle #1,\, #2 \right\rangle}
\def\dotlb[#1,#2]{\left\langle #1,\, #2 \right\rangle}
\def\dotlm[#1,#2]{\left[ #1,\, #2 \right]}
\def\dotp[#1,#2]{(\vect{#1} \cdot\vect{#2})}
\def\aff[#1,#2]{\hat{#1}(#2)}
\def\n4sym{{\cal N}=4 SYM}
\def\>{\rangle}
\def\<{\langle}
\def\weight[#1,#2,#3]{\{(#1),#2,#3\}}
\def\ads[#1]{$\text{AdS}_{#1}$}
\def\cft[#1]{$\text{CFT}_{#1}$}

\newcommand{\be}{\begin{equation}}
\newcommand{\ee}{\end{equation}}
\newcommand{\ba}{\begin{align}}
\newcommand{\ea}{\end{align}}
\newcommand{\bs}{\begin{split}}
\def\sess\end{split}

\newcommand{\vect}[1]{{\boldsymbol{#1}}}

\notoc
\title{The Sensitivity of Black Holes to Low Energy Excitations}
\author{Suvrat Raju}
\affiliation{International Centre for Theoretical Sciences, Tata Institute of Fundamental Research, Shivakote, Bengaluru 560089, India.}
\emailAdd{suvrat@icts.res.in}

\date{}

\abstract{A general principle of statistical mechanics is that low energy excitations of a thermal state change expectation values of observables by a very small amount. However some observables in the vicinity of the horizon of a large black hole in anti-de Sitter space naively seem to violate this bound. This potential violation is related to the question of whether the black hole interior can be described in AdS/CFT.  Here we point out that if the possible excitations are limited to those produced by a simple local source on the boundary, and the possible observables are limited to products of field operators in  a single causal patch in the bulk, then these violations disappear. 
\vskip 1in
{\center{\bf Essay written for the Gravity Research Foundation 2017 Awards for Essays on Gravitation.}
}
}
\begin{document}
\maketitle
In this essay we will describe a paradox that appears when a general result from statistical mechanics is applied to the geometry of a large AdS black hole. This paradox is closely related to the  
question of whether the interior of a large black hole can be described holographically. Furthermore, the resolution that we describe here sheds light on the question of how classical observables are represented in a theory of quantum gravity. So, although this paradox is not very well known, we believe that it deserves attention.

\paragraph{\bf A result from statistical mechanics \\}
Consider a typical state in a system with a large number of degrees of freedom. By typical state, we mean a linear combination of energy eigenstates with randomly chosen coefficients. It is well known that such a state,  with mean energy $E$,  resembles a thermal state at an inverse temperature $1/T = \beta = {\rho'(E)/\rho(E)}$, where $\rho(E)$ is the density of states at that energy.  Now consider exciting the system with a unitary operator that increases the energy by $\delta E \ll T$.   Then a general result from statistical mechanics states that this excitation changes the expectation value of any observable by only a small amount compared to the spontaneous fluctuations in the observable's value.

Intuitively, this result is simple to understand.  When the system is in equilibrium, its smallest constituent subsystem has an energy of order $T$. 
So an excitation with energy lower than $T$ is unlikely to have an easily measurable effect.  This intuition can be converted into a precise theorem \cite{Raju:2016vsu}. Consider a normal operator $\al$,  a typical state $|\Psi \rangle$ with energy $E$,  and a unitary $U$ that changes the energy by $\delta E$, i.e., if the Hamiltonian of the system is $H$ then $\langle \Psi | H | \Psi \rangle = E$ and $\langle \Psi | U^{\dagger} H U |\Psi \rangle = E + \delta E$. Then the change induced in the expectation value of $A$, by the excitation,  is bounded by
\be
\label{bound}
\delta  \al  \equiv \big|\langle \Psi | U^{\dagger} \al U |\Psi \rangle - \langle \Psi | \al |\Psi \rangle \big| \leq 2 \sqrt{\beta \delta E} \sigma,
\ee
in the limit where $\beta \delta E \ll 1$ and where 
\be
\label{sigmadef}
\sigma^2 \equiv \langle \Psi| \al^{\dagger} \al | \Psi \rangle   - \left| \langle \Psi | \al | \Psi \rangle \right|^2,
\ee
measures the fluctuations of the observable.

\begin{figure}[!h]
\begin{center}
\begin{subfigure}[t]{0.4\textwidth}
\begin{center}
\includegraphics[height=0.3\textheight]{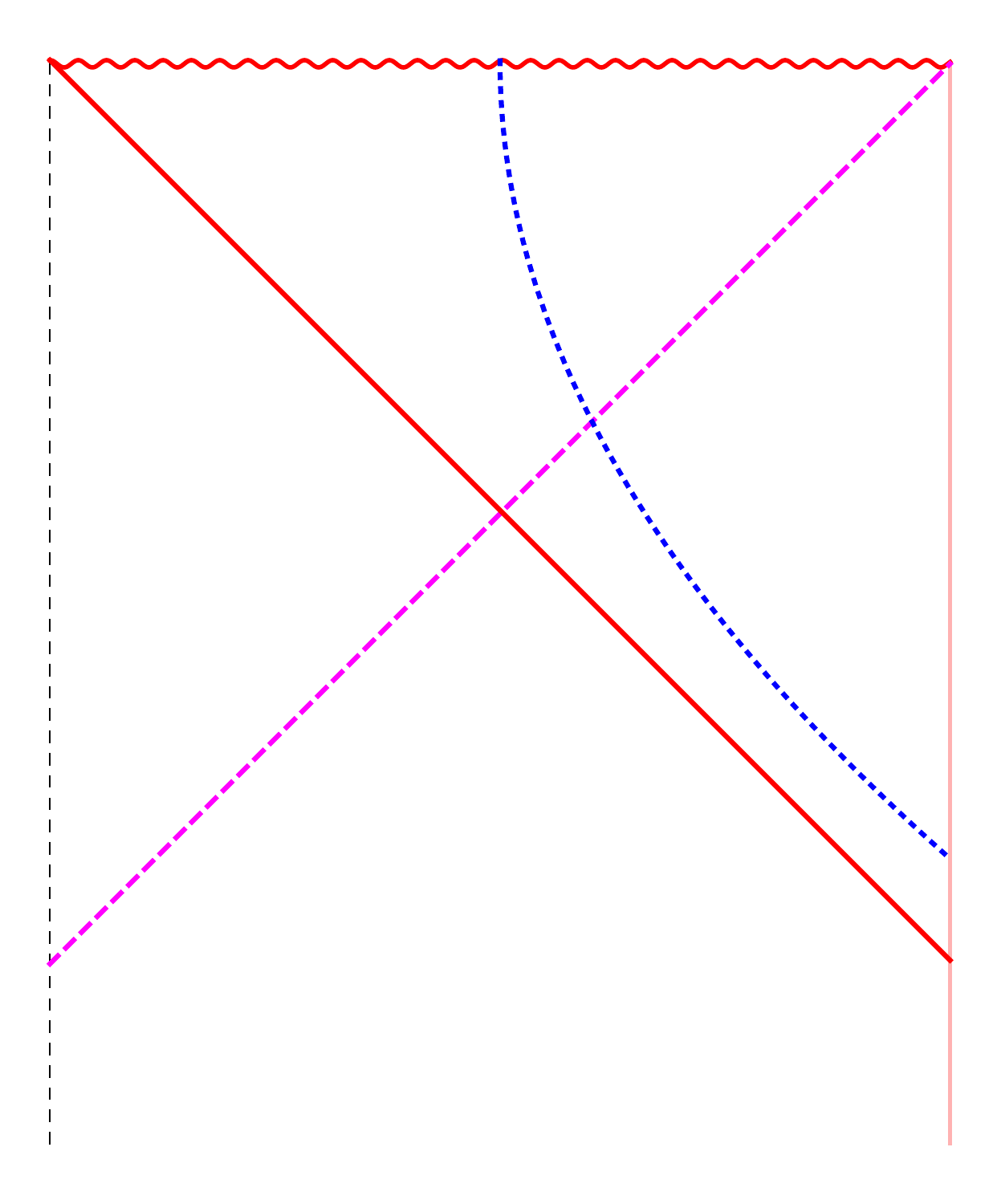}
\end{center}
\caption{\label{largeadsnotexcited}}
\end{subfigure}
\qquad \qquad \qquad
\begin{subfigure}[t]{0.4\textwidth}
\begin{center}
\includegraphics[height=0.3\textheight]{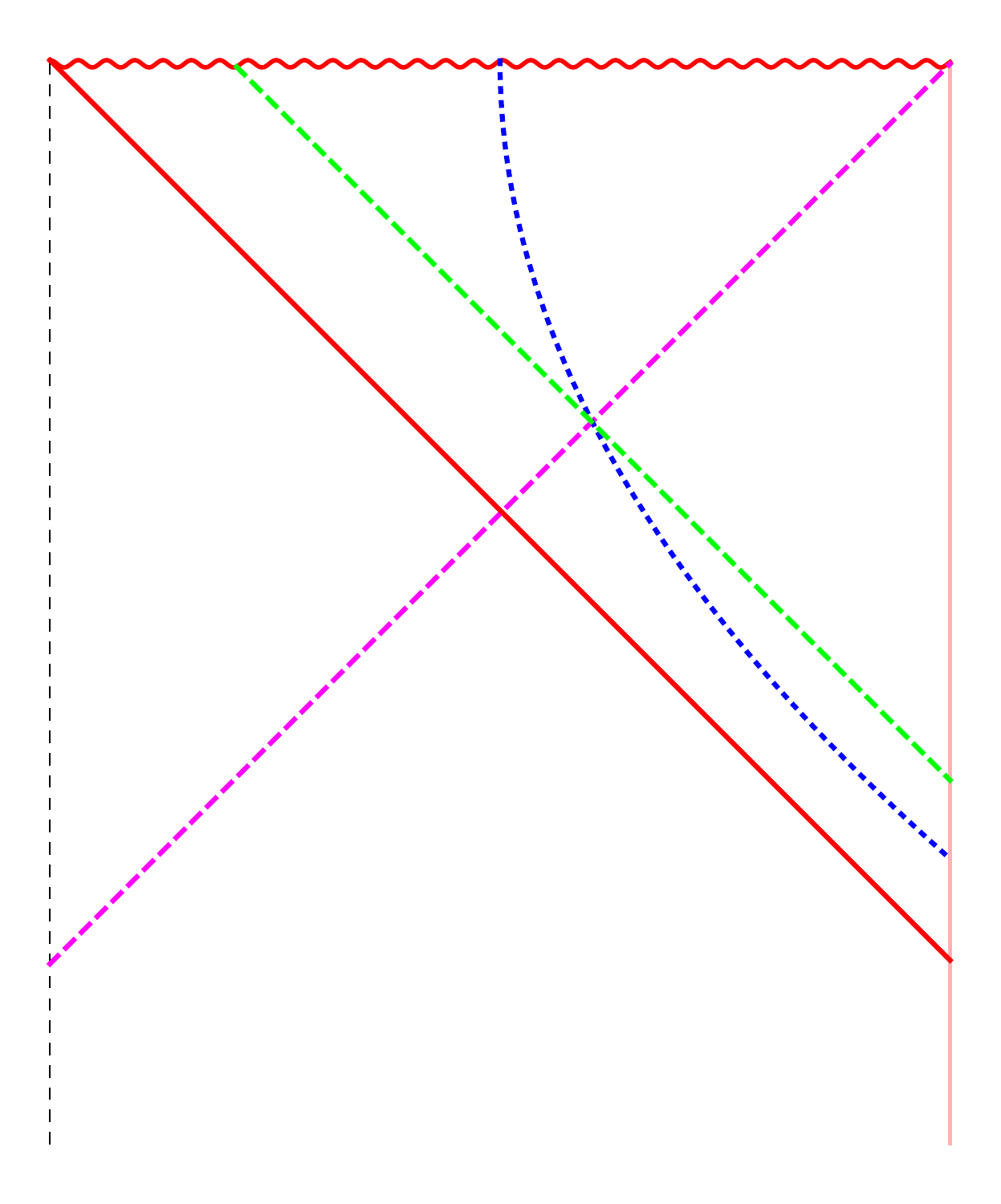}
\end{center}
\caption{\label{largeadsexcited}}
\end{subfigure}
\caption{\it On the left, we display the worldline of an infalling observer (dotted blue) in a black hole produced by a collapsing shell (solid red). On the right, a source produces an excitation (dashed green) that collides with the observer near the horizon.}
\end{center}
\end{figure}

\paragraph{\bf A paradox involving  large AdS black holes \\}
Consider a large AdS black hole formed from collapse, as shown in figure \ref{largeadsnotexcited}. The action of a unitary can change the state to the one shown in figure \ref{largeadsexcited} with an excitation near the horizon. The paradox results from the fact that, in a gravitational theory, the Hamiltonian of the system is  defined asymptotically near the boundary \cite{DeWitt:1967yk,Regge:1974zd,Balasubramanian:1999re}. So  $\delta E$, as measured by the asymptotic Hamiltonian, may be small even if the excitation has large local energy.  Now it seems that the excitation could change a cleverly designed local correlator, which  may be sensitive to its local energy, by a significant amount in violation of \eqref{bound}. This issue was first highlighted in \cite{Marolf:2015dia, Harlow:2014yoa} and then sharpened in \cite{Raju:2016vsu}.

\paragraph{\bf Relation to the the black hole interior \\}

It is not difficult to see that an observer who always stays outside the horizon is not in danger of observing any violation of \eqref{bound}. For example, consider an observer who hovers at a constant  height above the horizon. Compared to the boundary observer, the local temperature and energy seen by the hovering observer are both larger by a red-shift factor.  Since it is only the ratio of these quantities that enters the bound \eqref{bound}, if the bound is preserved in the frame of the hovering observer, it will also hold for the asymptotic observer. 

This intuition can be made precise as follows.  The AdS/CFT conjecture tells us that field operators outside the black hole are linear combinations of simple local operators in a holographic theory \cite{Hamilton:2006fh,Hamilton:2006az,Papadodimas:2012aq}.
Since these boundary operators obey \eqref{bound} in the boundary theory, we see that correlators of operators outside the black hole will also obey \eqref{bound}.

The situation is more delicate for the infalling observer who may see no temperature in her frame, but may still suffer a high-energy collision with the excitation if the relative boost between the two is large. Stated more technically,  an implicit assumption in the derivation of \eqref{bound} is that the operator $\al$ is state-independent.  However, \cite{Papadodimas:2015jra,Papadodimas:2015xma,Papadodimas:2013jku,Papadodimas:2013wnh} showed that if operators inside the black hole exist in the CFT, and if we wish to avoid a firewall at the horizon, then these interior operators must be state-dependent. For state-dependent operators, we need to check the bound \eqref{bound} by hand.

\paragraph{\bf A set of reasonable experiments \\}

The physically important question is whether the bound \eqref{bound} holds within the set of experiments that can be performed by a bulk observer with finite powers. This places restrictions on the allowed excitations and observables as we now describe.

To excite the geometry, we allow the observer to deform the Hamiltonian through
\be
\label{deformation}
H_{\text{deformed}}(t) =  H + J(t) S(t),
\ee
where $J(t)$ is a c-number source and $S(t)$ is an operator near the boundary of AdS at time $t$. We emphasize that, in quantum gravity, the set of gauge invariant operators is spanned by operators near the boundary and so asymptotic sources exhaust the set of allowed excitations.  This is equivalent to the colloquial observation that bulk sources in gravity would violate local energy conservation; one can only inject energy into the system by throwing it in from the boundary.

But, we need to place additional restrictions on the possible sources. In \cite{Papadodimas:2015jra,Papadodimas:2015xma,Papadodimas:2013jku,Papadodimas:2013wnh}, we excluded sources dual to very heavy, or complicated, operators.  The action of the remaining excitations led to an effective Hilbert space, ${\cal H}_{\Psi}$, that was sufficient to describe bulk effective field theory, 
and interior operators were constructed within this space.

\begin{figure}
\begin{center}
\includegraphics[height=0.5\textwidth]{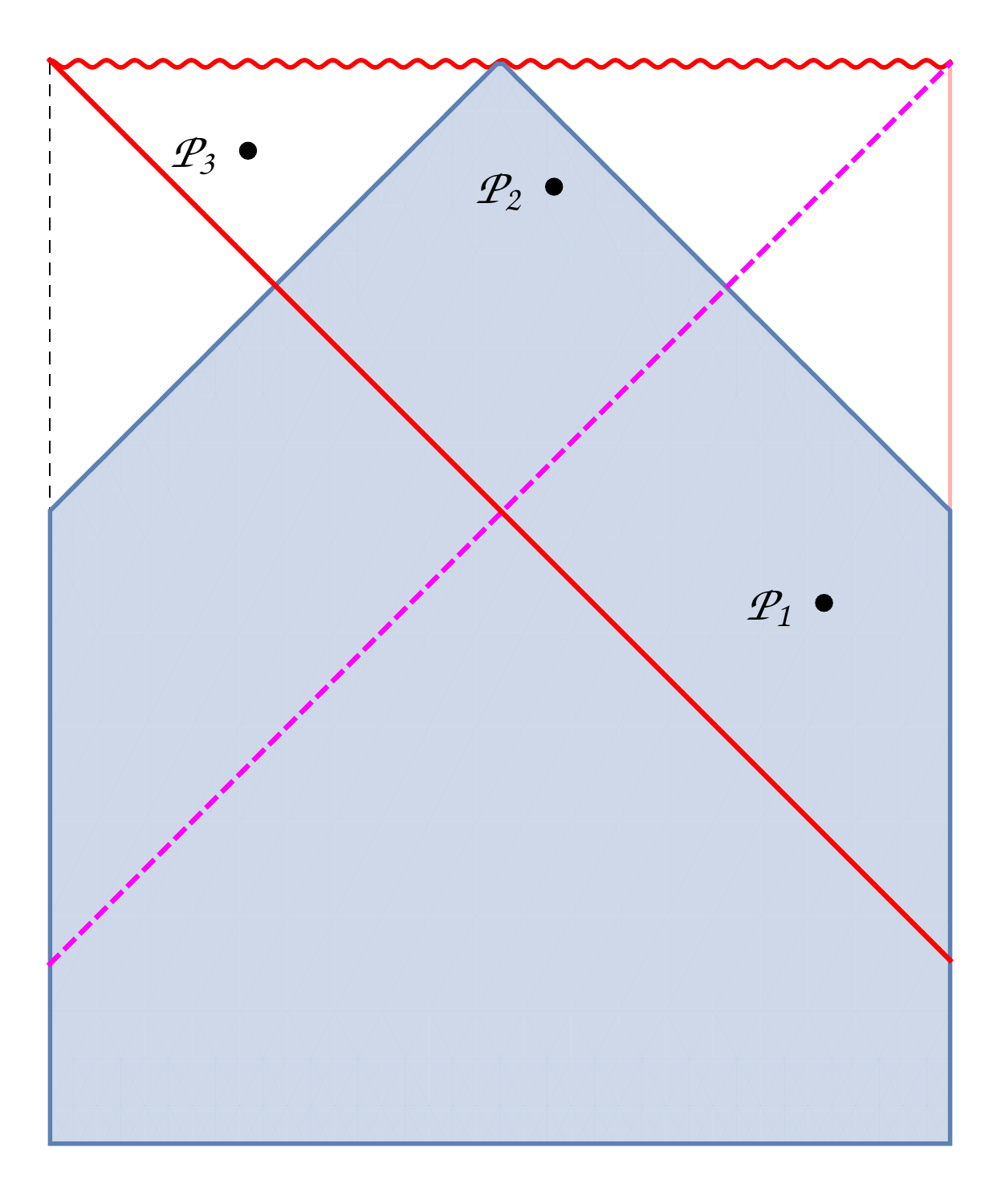}
\caption{\it A causal patch is the past null cone of some point on the singularity.  Points ${\cal P}_1$ and ${\cal P}_2$ are in a single patch; but the three points ${\cal P}_1, {\cal P}_2, {\cal P}_3$ do not fit into any patch. \label{causalingeon}}
\end{center}
\end{figure}

In this essay, we will impose {\em two} additional restrictions that were not imposed in \cite{Papadodimas:2015jra,Papadodimas:2015xma,Papadodimas:2013jku,Papadodimas:2013wnh}. 
\begin{enumerate}
\item[\hypertarget{r1const}{\bf R1}]
We will consider excitations, where  $S(t)$ is a simple boundary operator {\em localized} at \mbox{time $t$}.
\item[\bf R2\label{r2const}]
We will consider observables, $\al = \phi({\cal P}_1) \ldots \phi({\cal P}_n)$, that are products of local bulk field operators whose locations ${\cal P}_i$ fit in a {\em single causal patch} in the bulk. (See figure \ref{causalingeon}.)
\end{enumerate}
To understand restriction \hyperlink{r1const}{R1},  let $O(t)$ be a boundary operator dual to some bulk field. Then, in principle, a possible deformation of the Hamiltonian is  $H_{\text{deformed}}(t) = H +$  \mbox{$J(t) O(t + \tau)$}, where $\tau$ is some constant. But since $O$ does not obey an equation of motion,  $O(t + \tau)$ involves very  complicated operators when expressed in terms of operators at time $t$. 
So restriction  \hyperlink{r1const}{R1} disallows sources of this form, except when $\tau = 0$. 

This restriction is reasonable because an observer with finite powers can only turn on sources that couple to the instantaneous value of a field, as opposed to its past or future values. Indeed, it is easy to see that  an observer with the ability to turn on sources that couple to the future value of a field can engineer violations of the second law of thermodynamics near the boundary.

To justify restriction  \hyperlink{r2const}{R2}, we simply note that observables that do not fit in a single causal patch cannot be measured by any bulk observer.

\paragraph{\bf Consistency of the interior \\}
It turns out that within the class of experiments that obey the restrictions  \hyperlink{r1const}{R1} and  \hyperlink{r1const}{R2}, there are no violations of the bound \eqref{bound}! This was proved in \cite{Raju:2016vsu} using a surprising property of position-space AdS correlators.

The main idea of the proof is easy to state. For notational simplicity, we  take  $\al$ to be a product of operators that lie on a spacelike slice within the patch and refer the reader to \cite{Raju:2016vsu} for the general case.  We can choose coordinates so that this slice intersects the boundary on a surface of constant time, $t_0$.   Then the source \eqref{deformation} modifies the Heisenberg observable, $\al$, through $\al \rightarrow U^{\dagger} \al U$ where  $U =  {\cal T}\{e^{-i \int_{-\infty}^{t_0} d t\, J(t) S(t)}\}$ and  ${\cal T}$ is the time-ordering symbol. Note that the function $J(t)$ may have support for times larger than $t_0$ --- even all the way up to $t = \infty$ --- but this does not enter into expectation values calculated at $t_0$. 

Now field operators inside the black hole are state-dependent but \cite{Raju:2016vsu} showed that, under the restrictions above,  it is possible to find a {\em state-independent} operator, $\widehat{\al}$, that approximates $\al$ both in the presence and the absence of the source.
\be
\begin{split}
&\langle \Psi | \al | \Psi \rangle = \langle \Psi | \widehat{\al} | \Psi \rangle; \\
&\langle \Psi | U^{\dagger} \al U | \Psi \rangle  = \langle \Psi | U^{\dagger} \widehat{\al}  U | \Psi \rangle  + \Or[\beta \delta E].
\end{split}
\ee
As an ordinary operator in the boundary theory,  $\widehat{\al}$ must obey  $\delta \widehat{A} \leq 2 \sqrt{\beta \delta E} \,  \widehat{\sigma}$, and a careful consideration of its fluctuations shows that $\widehat{\sigma} = \sigma$. But these relations imply that even $\delta \al$ obeys the bound \eqref{bound}. 

Thus while operators in the black hole interior may seem to have unusual properties, these properties are invisible to an ordinary bulk observer.

\paragraph{\bf Outlook \\}
The paradox outlined above is yet another instance where in seeking to describe the black hole interior holographically, one encounters puzzles that involve simple expectations from statistical mechanics.
Large AdS black holes do not evaporate but these puzzles can be thought of as versions of the information paradox.  They can be uniformly resolved by the hypothesis that
classically well defined questions about the black hole interior map to quantum operators  that behave like ordinary observables in simple experiments, but have unusual properties like state-dependence when examined globally on the Hilbert space. It remains a tantalizing open problem to unify these ideas into a holistic understanding of measurements in quantum gravity.

\bibliographystyle{JHEP}
\bibliography{references}

\end{document}